\documentclass[11pt,a4paper]{article}
\usepackage{times}

\usepackage{pgfplots}
\usepackage{fullpage}

\usepackage[utf8]{inputenc} 
\usepackage[T1]{fontenc}    
\usepackage{hyperref}       
\usepackage{url}            
\usepackage{booktabs}       
\usepackage{amsfonts}       
\usepackage{nicefrac}       
\usepackage{microtype}      
\usepackage{lipsum}		
\usepackage{graphicx}
\usepackage{natbib}
\usepackage{doi}

\usepackage{url}
\usepackage{caption}
\usepackage{graphicx}
\usepackage{grffile}
\usepackage{amsmath}
\usepackage{booktabs}
\usepackage{subcaption}
\usepackage{wrapfig}
\usepackage{natbib}
\usepackage{nicefrac}
\usepackage{pgfplots}
\usepackage{amsmath}
\usepgfplotslibrary{units} 
\usepgfplotslibrary{fillbetween} 
\usepgfplotslibrary{colormaps} 
\pgfplotsset{width=6cm, height=7cm}
\pgfplotsset{compat=newest} 
\mathchardef\mhyphen="2D 

\newcommand{\np}{{\mathrm{NP}}}

\newcommand{\tfidf}{{{TF\hbox{-}IDF}}}

\newcommand{\tf}{{\emph{TF}}}
\newcommand{\idf}{{\emph{IDF}}}

\newcommand{\lambdascore}{\lambda\mhyphen score}
\newcommand{\lambdakscore}{\lambda^{(k)}\mhyphen score}
\newcommand{\utilityE}{\breve{E}}

\newcommand{\localapp}{{\mathit{local}}}
\newcommand{\local}{{{\mathit{util}}}}

\newcommand{\reals}{\mathbb{R}}

\newtheorem{example}{Example}

\newcommand{\rank}{{{\mathrm{rank}}}}
\newcommand{\dis}{{{\mathrm{diss}}}}
\newcommand{\diss}{{\dis}}
\newcommand{\ext}{{{\mathrm{ext}}}}

\title{Using Multiwinner Voting to Search for Movies}
\date{\today}

\author{ 
    {Grzegorz Gawron} \\
  AGH University and VirtusLab\\
  Krakow, Poland\\
  {\small\texttt{ggawron@virtuslab.com}} \\
  \and
    {Piotr Faliszewski} \\
  AGH University\\
  Krakow, Poland\\
  {\small \texttt{faliszew@agh.edu.pl}} \\
}

\begin{document}

\maketitle

\begin{abstract}  
  We show a prototype of a system that uses multiwinner voting to
  suggest resources (such as movies) related to a given query set
  (such as a movie that one enjoys). Depending on the voting rule used,
  the system can either provide resources very closely related to the
  query set or a broader spectrum of options. We show how this ability can be
  interpreted as a way of controlling the diversity of the results. We
  test our system both on synthetic data and on the real-life
  collection of movie ratings from the MovieLens dataset. We also
  present a visual comparison of the search results corresponding to
  selected diversity levels.
\end{abstract}

\section{Introduction}

The idea of multiwinner voting is to provide a committee of candidates
based on the preferences of the voters. In principle, such mechanisms
have many applications, ranging from choosing parliaments, through selecting finalists of
competitions, to suggesting items in Internet stores or
services. While the first two types of applications indeed are quite
common in practice, the last one, so far, was viewed mostly as a
theoretical possibility. Our goal is to change this view.  To this
end, we design a prototype of a voting-based search system that given a
movie (or, a set of movies), finds related ones. The crucial element of our system---enabled by the
use of multiwinner voting---is that one may specify to what extent he
or she wants to focus on movies very tightly related to the given one,
and to what extent he or she wants to explore a broader spectrum of
movies that are related in some less obvious ways.
Indeed, if someone is looking for movies exactly like the specified
one, then using focused search is natural. However, if someone is not
really sure what he or she really seeks, or if he or she has already
watched the most related movies, looking at a broader spectrum is more
desirable.

Viewed more formally, our system belongs to the class of
non-personalized recommendation systems based on collaborative
filtering. That is, from our point of view the users posing queries
are anonymous and we do not target the results toward particular
individuals, but rather we try to find movies related to the ones they
ask about. In this sense, we provide more of a search-support tool
than a classical recommendation system.

To find the relationships between the movies, we use a dataset of
movie ratings (in our case, the MovieLens dataset of
\citet{har-kon:j:movielens}. Such a dataset consists of a set of
agents who rate the movies on a scale between one and five stars
(where one is the lowest score and five is the highest). For each
movie we consider which agents enjoyed it and what other movies these
agents liked. More specifically, 
given the raw data with movie ratings we form a global election where
we indicate which users liked which movies (we say that a user liked a
movie if he or she gave it at least four stars; in the language of
voting literature, liking a movie corresponds to \emph{approving} it).
Then, given a query, i.e., either a single movie or a set of movies,
we restrict this election to the agents who liked the movies from the
query (and the movies they liked, except for the ones from the query).
Based on this local election, for each user and each movie that he or
she likes, we determine a utility score which indicates how relevant
the movie is (briefly put, we need to distinguish between movies that
are globally very popular, such as, e.g., \emph{The Lord of the
  Rings}, from the ones that are mostly popular among the agents in
the local election). Finally, we seek a winning committee with respect
to one of the OWA-based multiwinner voting rules discussed by
\citet{sko-fal-lan:j:collective}, and output its contents as our
result (see also the works of
\citet{azi-bri-con-elk-fre-wal:j:justified-representation} and
\citet{bre-fal-kac-kno-nie:c:collective-parameterized} for further
discussions of these rules). Since, in general, our rules are
$\np$-hard to compute, we use approximation algorithms and heuristics.

OWA-based rules are parameterized by the \emph{ordered weighted
  average} operators of \citet{yag:j:owa} and, depending on the choice
of these operators, they may provide committees of very similar,
individually excellent candidates, or of more diverse ones. This is
illustrated in the simulations of
\citet{elk-fal-las-sko-sli-tal:c:2d-multiwinner},
\citet{fal-sko-sli-tal:c:paths}, and
\citet{god-bat-sko-fal:c:2d-approval}, and explained theoretically by
\citet{azi-bri-con-elk-fre-wal:j:justified-representation} and
\citet{lac-sko:c:approval-thiele,lac-sko:j:quantitative-mw}).  Thus,
by choosing the OWA operators appropriately, we either find movies
very closely related to a given query, or those that form a broader
spectrum of related movies. Specifically, we use a family of operators
parameterized by a value $p \geq 0$, such that for $p=0$ we get the
most focused results, and for larger $p$'s the results become more
broad.

\paragraph{Our Contribution.}
Our main contribution is designing a voting-based search system and
testing it in the context of selecting movies. In particular, we show
the following results:
\begin{enumerate}
\item Using movie-inspired synthetic data, we show that, indeed, the
  rules that are meant to choose closely related movies or more broad
  committees, do so.  Using the MovieLens dataset, observe that the
  system provides appealing results on sample queries.

\item We use our system  to visualize relations
  between movies. On the one hand, this allows us to observe that our voting rules
  act as intended also on real-life data. On the other hand, it provides insight
  into the nature of the movies. 
  
\item We show that the greedy algorithm is, on average, superior to
  simulated annealing for computing our committees (but simulated
  annealing also has positive features).

\end{enumerate}
While our system is a prototype, we believe that its results are
promising and deserve further study.

\paragraph{Related Work.}
Regarding multiwinner voting, we point the readers to the overview of
\citet{fal-sko-sli-tal:b:multiwinner-voting}, who give a broad
overview of multiwinner voting rules and their possible applications,
and to the survey of \citet{lac-sko:t:approval-multiwinner-survey},
who focus on approval-based rules. It is also interesting to consider
the work of \citet{elk-fal-sko-sli:j:multiwinner-properties}, where
the idea of using multiwinner voting for selecting movies was
suggested (albeit, in a somewhat different setting).
While multiwinner voting is not yet a mainstream tool in applications,
various researchers have used it successfully.  For example,
\citet{cha-pat-gan-gum-loi:c:fairness-rec-sys} have shown that
appropriate multiwinner rules can be used to select trending topics on
Twitter or popular news on the Internet.  For the latter task,
\citet{mon-bal-sin-pat:c:rec-sys-fairness} also designed a
voting-based solution.
\citet{pou-kel-esh:c:multiwinner-genetic} and
\citet{fal-saw-sch-smo:j:genetic-multiwinner} used diversity-oriented
multiwinner voting rules to design genetic algorithms that would avoid
getting stuck in a plateau regions of the search space (i.e., in the
areas where the objective function has nearly constant value; when a
genetic algorithm identifies such an area, it is beneficial if it
explores its shape, instead of getting stuck in a local optimum).

For a broad discussion of modern recommendation systems, we point to
the handbook edited by \citet{ric-rok-sha:b:rec-sys}.  For an early
account of collaborative filtering methods, we point to the work of
\citet{sar-kar-kon-rie:c:movies-rec-sys}.  As examples of works on
movie recommendations, we mention the paper of
\citet{gho:j:rec-sys-movie}, who describes a movie recommendation
system using Black's voting rule with weighted user preferences, the
paper of \citet{aza-has-kra-esh-wei-nat:c:movie-rec-sys}, who focus on
maximizing the revenue of the recommender, the paper of
\citet{cho-ko-han:j:rec-sys-movies}, who discuss recommendations based
on movie genres, and the paper of
\citet{pho-par-xin-hao:j:rec-sys-movie}, who adapt some techniques
from social networks to recommendation systems.  While most of this
literature considers the most tightly connected movies---and as such
is less relevant to our study---there are works on recommendation
systems that focus on diversifying the results; see, e.g., the work of
\citet{kim-kim-par-yu:c:rec-sys-diverse} (the main difference between
their work and ours is that they use neural networks and rely on a
number of features, whereas we use multiwinner voting and provide a
solution based on simple collaborative filtering).
Thus, so far, we have not found studies whose results we could
directly compare to ours, except, of course, for those looking for the
most related movies. However, the main point of our study is to
provide diverse results, for situations where the user him or herself
does not know what he or she is looking for. In this sense, our work
can be seen as providing a way to break information bubble in which
the user might have locked him or herself.

Throughout the rest of this section, let us review notions of
diversity that appear in the context of voting, recommendation
systems, and information retrieval.  \citet{drosou2017diversity}
pointed out that diversity is an ambiguous notion and is understood
differently, depending on the context.  We are committed to the
multiwinner elections context, as presented by
\citet{faliszewski2017multiwinner}, where a diverse winning committee
is expected to represent as many voters as possible (formally, this
can be understood as maximizing the number of voters that approve at
least one member of the committee).  This approach is similar to the
concept of representativeness described by
\citet{Chasalow2021RepresentativenessIS}, where the problem is to
select a subsample that \emph{represents} a larger population.
\citet{celis2017multiwinner} additionally seek committees whose
members are diverse in terms of additional features---such as, e.g.,
gender or seniority level. We do not follow them in this respect as in
our model we do not have access to movies' features.

Next we discuss other notions of diversity, which are orthogonal to
our approach, but whose knowledge helps in understanding our scope.
\citet{clarke2008novelty} describe diversity as a tool to respond to
user's potential multiple intents when using an ambiguous query.  This
is different from our approach because we do not expect the user to
has some specific intent that we do not know, but rather that the user
is not sure him or herself what the intent truly is.
\citet{clarke2008novelty} also define a related concept of
\emph{novelty} to avoid returning duplicate items if present, but this
does not apply to our setting as we provide non-personalized search,
which does not keep track of the history of the queries.
\citet{mitchell2020diversity} refer to the variety in the result set
with respect to any potential candidate characteristic as
\emph{heterogeneity}, while reserving the term \emph{diversity} to
variety with respect to purely sociopolitical characteristics, such as
gender, age or race.  Yet another viewpoint is presented by
\citet{celis2016fair}, who define \emph{combinatorial diversity} to
mean the entropy of the distribution of candidate features in the
result set and \emph{geometric diversity} to relate to the volume of a
hyper cube with vertices being the space-embedded candidates.  We find
it hard to use the standard measures of diversity used in the
recommendation and information retrieval research, such as nDCG, ERR,
ERR-IA, AP or bpref (for definitions see \citep{leung2018encyclopedia}
and \citep{jarvelin2002cumulated}). Indeed, all of them depend on
defined order of results or/and defined query intents.

\section{Preliminaries}\label{sec:preliminaries}\label{sec:prelim-using}

Let $\reals_+$ denote the set of nonnegative real numbers, and for a
positive integer $i$, let $[i]$ denote the set $\{1, \ldots, i\}$.

\paragraph{Utility and Approval Elections.}
Let $R = \{r_1, \ldots, r_m\}$ be a set of \emph{resources} and let
$N = \{1, \ldots, n\}$ be a set of \emph{agents} (in other papers, the
resources are often referred to as the \emph{candidates} and the
agents are often referred to as the \emph{voters}). Each agent $i$ has
a utility function $u_i \colon R \rightarrow \reals_+$, which
specifies how much he or she appreciates each resource. We assume that
the utilities are comparable among the agents and that the utility of
zero means that an agent is completely uninterested in a given
resource. We do not normalize utility values, so, for example, some
agent may be far more excited about the resources than some other
one. Committees are sets of resources, typically of a given size~$k$.
For a committee $S = \{s_1, \ldots, s_k\}$ and an agent $i$, by
$u_i(S)$ we mean the vector $(u_i(s_1), \ldots, u_i(s_t))$, where the
utilities appear in some fixed order over the resources (this order
will never be relevant for our discussion).
We write $U = (u_1, \ldots, u_n)$ to denote a collection of utility
functions, referred to as a \emph{utility profile}.
A \emph{utility election} $E = (R,U)$ consists of a set of resources
and a utility profile over these resources (a utility profile
implicitly specifies the set of agents).
An \emph{approval election} is a utility election where each utility
is either $1$, meaning that an agent approves a resource, or $0$,
meaning that he or she does not approve it.  For approval elections we
typically denote the utility profile as $A = (a_1, \ldots, a_n)$ and
call it an \emph{approval profile}.
For a resource $r_t$, we write $A(r_t)$ to denote the set of agents that
approve it.

\paragraph{OWA Operators.}
An \emph{ordered weighted average} (OWA) operator is 
specified by a vector of nonnegative real numbers, such as
$\lambda = (\lambda_1, \ldots, \lambda_k)$, and operates as follows.
For a vector $x = (x_1, \ldots, x_k) \in \reals^k$
and a vector $x' = (x'_1, \ldots, x'_k)$ obtained by sorting $x$
in the nonincreasing order, we have:
\[
  \lambda(x) = \lambda_1 x'_1 + \lambda_2 x'_2 + \cdots + \lambda_k
  x'_k.
\]
For example, operator $(1, \ldots, 1)$ means summing up the elements
of the input vector, whereas operator $(1,0,\ldots, 0)$ means taking
its maximum element. OWA operators were introduced by
\citet{yag:j:owa}.

\paragraph{(OWA-Based) Multiwinner Voting Rules.}
A \emph{multiwinner voting rule} is a function $f$ which, given a
utility election $E$ and an integer~$k$, returns a family of size-$k$
winning committees.  We focus on OWA-based rules.

Consider a utility election $E = (R,U)$, where
$R = \{r_1, \ldots, r_m\}$ and $U = (u_1, \ldots, u_n)$, and an OWA
operator $\lambda = (\lambda_1, \ldots, \lambda_k)$. Let $S$ be a
size-$k$ committee.  We define the $\lambda$-score of committee $S$
in election $E$ to be
$\lambdascore_E(S) = \sum_{i=1}^{n} \lambda(u_i(S))$.  We say that a
multiwinner rule $f$ is OWA-based if there is a family
$\Lambda = (\lambda^{(k)})_{k \geq 1}$ of OWA operators, one for each
committee size~$k$, such that for each election $E$ and each committee
size~$k$, $f(E,k)$ consists exactly of those size-$k$ committees $S$
for which $\lambdakscore_E(S)$ is highest.

\paragraph{HUV Rules.}
We are particularly interested in the rules that use OWA operators of the
followig form, where $p \geq 0$:
\[
  \lambda^p = (1, \nicefrac{1}{2^p}, \nicefrac{1}{3^p}, \ldots ),
\]
and we refer to them as $p$-\emph{Harmonic Utility Voting} rules
($p$-HUV rules). The name stems from the fact that for $p = 1$, their
OWA operators sum up to harmonic numbers.  Let us consider three
special cases:
\begin{enumerate}
\item For a committee size $k$, the $0$-HUV rule chooses $k$ resources
  with the highest total utility;

  indeed, its OWA operator is $(1, \ldots, 1)$. Under approval
  elections, $0$-HUV is the classic \emph{Multiwinner Approval Voting}
  rule (AV).
\item The $1$-HUV rule uses OWA operators
  $(1, \nicefrac{1}{2}, \nicefrac{1}{3}, \ldots)$; for approval
  elections this is the \emph{Proportional Approval Voting} rule (PAV) of
  \citet{thi:j:pav}.
\item Abusing the notation, $\infty$-HUV is a rule that uses OWA
  operators $(1,0, \ldots, 0)$; for approval elections it is the
  \emph{Chamberlin--Courant} rule (CC); originally introduced by
  \citet{cha-cou:j:cc} for the ordinal setting and converted to the approval
  one by
  \citet{pro-ros-zoh:j:proportional-representation} and
  \citet{bet-sli-uhl:j:mon-cc}.
\end{enumerate}
In the approval voting setting, these three rules correspond to the
three main principles of choosing committees. AV chooses
\emph{individually excellent} resources, i.e., those that are
appreciated by the largest number of agents; PAV chooses committees
that, in a certain formal sense, proportionally represent the
preferences of the
agents~\citep{azi-bri-con-elk-fre-wal:j:justified-representation,bri-las-sko:j:multiwinner-apportionment},
and CC ($\infty$-HUV) focuses on \emph{diversity}, i.e., it seeks a
committee so that as many agents as possible appreciate at least one
item in the committee. For a more detailed description of these
principles, see the overview of
\citet{fal-sko-sli-tal:b:multiwinner-voting}. For a focus on approval
rules, see the survey of \citet{lac-sko:t:approval-multiwinner-survey}
and their work on the opposition between AV and CC
\citep{lac-sko:j:quantitative-mw}.

We proceed under two premises. The first one is that the $0$-HUV,
$1$-HUV, and $\infty$-HUV rules extend the principles of individual
excellence, proportionality, and diversity to the setting of utility
elections (the visualizations of
\citet{elk-fal-las-sko-sli-tal:c:2d-multiwinner} and \citet{god-bat-sko-fal:c:2d-approval} support this
view). The second one is that for $p > 1$, the rule $p$-HUV provides
committees that achieve various levels of compromise between those of
$1$-HUV and $\infty$-HUV (this is supported by the results of
\citet{fal-sko-sli-tal:c:paths}. We do not consider $p$ values between
$0$ and $1$).

\paragraph{Computing HUV Committees.}
Unfortunately, for each $p > 0$ it is $\np$-hard to tell if there is a
committee with at least a given score under the $p$-HUV
rule~\citep{sko-fal-lan:j:collective,azi-gas-gud-mac-mat-wal:c:approval-multiwinner}
and, as a consequence, no polynomial-time algorithms are known for
these rules (for $0$-HUV it suffices to sort the candidates in terms
of their total utilities and, up to tie-breaking, take top $k$
ones). We consider two ways of circumventing this issue:
\begin{enumerate}
\item We use the standard greedy algorithm: To compute a $p$-HUV
  committee of size-$k$ (for some $p > 0$), we start with an empty
  committee and perform $k$ iterations, where in each iteration we
  extend the committee with a single resource that maximizes its
  $p$-HUV score. A classic result on submodular optimization shows
  that the committees computed this way are guaranteed to achieve at
  least $1 - \nicefrac{1}{e} \approx 0.63$ fraction of the highest possible
  score~\citep{nem-wol-fis:j:submodular}.
  
\item We use the simulated annealing heuristic, as implemented in the
  \emph{simanneal} library, version 0.5.0. We set the number of steps
  to 50 000 and the temperature to vary between $9900$ and $0.6$.

\end{enumerate}
In principle, we could also use the formulations of $p$-HUV rules as
integer linear programs (ILPs), provided, e.g., by
\citet{sko-fal-lan:j:collective} and \citet{pet-lac:j:spoc}.  Yet,
given the sizes of our elections this would be quite infeasible (e.g.,
for the movie \emph{Alice in Wonderland (1951)} we obtain an election
with 32'783 resources and 5'339 agents).

\section{System Design}\label{sec:design}\label{sec:exploratory}
Let us now describe our voting-based search system. The main idea is
that for a \emph{query set} of resources (such as a set of movies that
someone enjoys) we form an election that regards related resources and
whose winning committee is our \emph{result}. Depending on the voting
rule used, the result may contain resources either very closely or only somewhat
loosely connected to those from the query.

The system consists of three main components, the data model, the
search model, and winner determination.

\subsection{Data Model}
The data model is responsible for converting domain-specific raw data
into what we call a \emph{global (approval) election}. For example,
the raw data may consist of information how various people rate
movies, or what products they buy in some store, or it may be
generated using some statistical model of preferences (which, indeed,
we will do to test our system in a controlled environment).

The global election stores our full knowledge of the domain. The
interpretation is that the agents are the users who have interacted
with some resources and they approve those for which the interaction
was positive (for example, if they enjoyed a particular movie).  Lack
of an approval either means that the interaction was negative or that
there was no interaction (while we could distinguish these two cases,
we find using basic approval elections to be simpler).

The algorithm for forming the global election is the only
domain-specific part of our model. Below we provide an example how
such an algorithm may work.

\begin{example}\label{ex:movielens}
  Consider the MovieLens 25M dataset~\citep{har-kon:j:movielens}. It
  contains 25'000'095 ratings of 62'423 movies, provided by 162'541
  users (so, on average, each user rated almost 154 movies).  Each
  rating is on the scale from one to five stars (the higher, the
  better) and was provided between January of 1995 and November of
  2019 on the MovieLens website. We form a global election where each
  user is an agent, each movie is a resource, and a user approves a
  movie if he or she gave it at least four stars.  We remove from
  consideration those movies that were approved by fewer than
  20~agents.

\end{example}

\subsection{Search Model}
The search model is responsible for forming a \emph{local (utility)
  election}, specific to a particular query. The idea is that this
election's winning committees would form our result sets.  We first
form a local approval election and then, if desired, we derive more
fine-grained utilities for the agents.

Let $E = (R,A)$ be the global approval election, where
$A = (a_1, \ldots, a_n)$, and let $Q \subseteq R$ be the query
set. Let $N = \{1, \ldots, n\}$ be the set of agents present in~$E$
and let~$N_\localapp$ be a subset of $N$ containing those agents who approve
at least one member of~$Q$ (while we could use other criteria, such as choosing agents that approve all members of $Q$, we
focus on singleton query sets for which it is not relevant). Then, let
$R_\localapp$ consist of those resources that are approved by the agents from
$N_\localapp$, except those from the query:
\[
  R_\localapp = \{ r \in R \mid (\exists i \in N_\localapp)[ a_i(r) = 1]\} \setminus Q.  
\]
Finally, let $A_\localapp$ be the approval profile of the agents from
$N_\localapp$, restricted to the resources from $R_\localapp$, and let
$E_\localapp = (R_\localapp, A_\localapp)$ be our local approval
election. Intuitively, it contains the knowledge about exactly those
resources that were appealing to (some of) the agents that also
enjoyed members of~$Q$.  Unfortunately, as shown below, it may be
insufficient to provide relevant search results.

\begin{example}\label{ex:needtfidf}
  Consider the MovieLens global election from
  Example~\ref{ex:movielens} and let the query set~$Q$ consist of a
  single movie, \emph{Hot Shots!}  (a 1991 parody of the \emph{Top Gun
    (1986)} movie, full of quirky/absurd humor). The five
  most-approved movies in the local approval election for~$Q$ are:
     (1)~\emph{The Matrix (1999)},
     (2)~\emph{Back to the Future (1985)},
     (3)~\emph{Fight Club (1999)},
     (4)~\emph{Pulp Fiction (1994)}, and
     (5)~\emph{Lord of the Rings: The Fellowship of the Ring (2001).}
  This is also the winning committee under the AV rule with
  $k=5$. Neither of these movies has much to do with \emph{Hot
    Shots!}, and they were selected because they are globally very
  popular (indeed, we expect many more readers of this paper to have
  heard of these five movies than of the one from the query set). Such
  globally popular movies are also popular among people enjoying
  \emph{Hot Shots!.}
\end{example}

To address the above issue, we derive a local utility election
$E_\local = (R_\local,U_\local)$, where $R_\local = R_\localapp$,
which promotes those resources that are more particular to a given
query.  To this end, we use the \emph{term frequency-inverse document
  frequency (TF-IDF)} mechanism.

\paragraph{TF-IDF.}

This is a standard heuristic introduced by
\citet{jon:j:tfidf2,jon:j:tfidf} to evaluate how specific is a given
term $t$ for a document $d$ from a document corpus $D$ (for further
information on TF-IDF see, e.g., the works of
\citet{robertson1997relevance} and~\citet{Ounis2018}).  The main idea
is that the specificity value of $t$ in $d$ is proportional to the
frequency of $t$ in $d$ (term frequency; \tf{}) and inversely
proportional to the frequency of $t$ in all the documents $D$ (inverse
document frequency; \idf{}).

Given our global election $E = (R,A)$ and the approval local election
$E_\localapp = (R_\localapp,A_\localapp)$, 
we implement the TF-IDF idea as follows. We interpret the resources as
the terms, and we take the document corpus to consist of two
``documents,'' election $E_\localapp$ and election $E' = (R_\localapp,A')$, where
$A'$ is the approval profile for those agents from the global
election that do not appear in $E_\localapp$.
Let $n$ be the total number of agents. For a resource $r \in R_\localapp$, we
let its \emph{term frequency} component be the number of agents that
approve it in the local election, i.e.,
$
  \mathrm{tf}(r) = |A_\localapp(r)|.
$
We let $r$'s inverse document frequency be
$
  \mathrm{idf}(r) = \ln\left( \nicefrac{n}{|A(r)|} \right).
$
Finally, to balance the TF and IDF components, we assume we have some
constant~$\gamma$ and we define:
\[
  \mathrm{tf}\hbox{-}\mathrm{idf}_\gamma(r) =
  \mathrm{tf}(r) \gamma^{\mathrm{idf}(r)} =
  \textstyle\frac{|A_\localapp(r)|}{|A(r)|^{\ln \gamma}} \cdot \left(n^{\ln \gamma} \right).
\]
\begin{example}
  Consider three resources, $r_1$, $r_2$, and $r_3$, where:
  $|A_\localapp(r_1)| = 1$, $|A(r_1)| = 2$, $|A_\localapp(r_2)| = 10$,
  $|A(r_2)| = 20$, and $|A_\localapp(r_3)| = 100$, $|A(r_3)| = 2000$.
  If we focused on the number of approvals in the local election (by
  taking $\ln\gamma = 0$), then we would view $r_3$ as the most
  relevant resource. 
  This would be unintuitive as only a small fraction of $r_3$'s
  approvals come from the agents who enjoy the items in the query
  set. For $ \gamma \approx 2.78$ (i.e., $\ln \gamma = 1$), we would
  focus on the ratios $\nicefrac{|A_\localapp(r_i)|}{|A(r_i)|}$, so $r_1$ and
  $r_2$ would be equally relevant, and $r_3$ would come third. This is
  more appealing, but still unsatisfying as, generally, $r_2$ is more
  popular than $r_1$.  By taking, e.g., $\gamma = 2$ (i.e.,
  $\ln \gamma \approx 0.69$) we would focus on ratios
  $\nicefrac{|A_\localapp(r_i)|}{|A(r_i)|^{0.69}}$ and, indeed, $r_2$ would be
  the most relevant resource, followed by~$r_1$ and~$r_2$.
\end{example}

We have found that $\gamma = 2$ works best for our scenario (we
discuss the process of choosing this value in
Section~\ref{sec:startrek}).

\begin{example}\label{ex:tfidf}
  Consider the same setting as in Example~\ref{ex:needtfidf}, but take
  five movies with the highest TF-IDF value (for $\gamma = 2$). We
  obtain:
  (1)~\emph{The Naked Gun 2 1/2 (1991)},
  (2)~\emph{Hot Shots! Part Deux (1993)},
  (3)~\emph{Top Secret! (1984)},
  (4)~\emph{The Naked Gun (1988)},
  (5)~\emph{The Loaded Weapon 1 (1993)}.
  All these movies are parodies similar in style to \emph{Hot Shots!}.
\end{example}

\paragraph{Local Utility Election.}
We form the local utility election $E_\local = (R_\local,U_\local)$ by
setting the utilities as follows. Given an agent $i$ from the local
approval election and a resource~$r \in R_\local$, if agent~$i$
approves~$r$, then set
$u_i(r) =
\nicefrac{\mathrm{tf}\hbox{-}\mathrm{idf}(r)}{|A_\localapp(r)|}$.
Otherwise, set it to be $0$. This way the utilities assigned to a
given resource sum up to its TF-IDF value.

\begin{example}
  By the design of the local utility election, a $0$-HUV committee of size
  $k=5$ for the \emph{Hot Shots!}  local utility election would
  consist exactly of the five movies listed in Example~\ref{ex:tfidf}.
\end{example}

\subsection{Winner Determination}
The last component of our system is to compute (an approximation of) a
winning committee under the local utility election under a given
$p$-HUV rule. If we are looking for resources that are the most
closely connected to the query set, then we take $p = 0$. For a
broader search, we consider $p \in \{1,2,3, \ldots\}$. Since
most of our rules are $\np$-hard to compute, we 
either use the greedy approximation algorithm or simulated
annealing. The greedy algorithm returns the committee ordered with
respect to the iteration number in which a given resource was added
(thus, the first resource is always the same for a given election,
irrespective of $p$). The algorithm based on simulated annealing
outputs the committee in an arbitrary order.

\begin{table*}
  \centering

  \resizebox{\textwidth}{!}{%
  \setlength{\tabcolsep}{5pt}
  \begin{tabular}{r|l|ll|ll}
    \multicolumn{1}{c|}{} & \multicolumn{1}{c|}{exact algorithm} & \multicolumn{2}{|c|}{simulated annealing} & \multicolumn{2}{|c}{greedy algorithm}\\
    \toprule
    \#\ \ & 0-HUV & 1-HUV & 2-HUV  & 1-HUV & 2-HUV\\
    \midrule
    1. & The Naked Gun 2 1/2 (1991)  & Hot Shots! Part Deux (1993)          & Hot Shots! Part Deux (1993) & The Naked Gun 2 1/2 (1991)   & The Naked Gun 2 1/2 (1991)     \\
    2. & Hot Shots! Part Deux (1993) & The Loaded Weapon 1 (1993)           & The Loaded Weapon 1 (1993)  & Hot Shots! Part Deux (1993)  & The Loaded Weapon 1 (1993) \\
    3. & Top Secret (1984)           & The Naked Gun 2 1/2 (1991)           & The Villain (1979)          & The Loaded Weapon 1 (1993)   & Major League II (1994)  \\
    4. & The Naked Gun (1988)        & Cannonball Run II (1984)             & Top Secret (1984)           & Major League II (1994)       & Yamakasi (2001) \\ 
    5. & The Loaded Weapon 1 (1993)  & Top Secret (1984)                    & Ernest Goes to Jail (1990)  & Top Secret (1984)            & Hot Shots! Part Deux (1993) \\
    6. & Police Academy (1984)       & Nothing to Lose (1997)               & Last Boy Scout, The (1991)  & Yamakasi (2001)              & To Be or Not to Be (1983)  \\
    7. & The Last Boy Scout (1991)   & Dragnet (1987)                       & Dragnet x(1987)             & Hudson Hawk (1991)           & Hudson Hawk (1991) \\
    8. & Commando (1985)             & Major League II (1994)               & Freaked (1993)              & To Be or Not to Be (1983)    & Freaked (1993)  \\
    9. & Hudson Hawk (1991)          & Yamakasi (2001)                      & Major League II (1994)      & City of Violence (2006)      & Top Secret (1984)  \\
    10.& Twins (1988)                & Coffee Town (2013)                   & Yamakasi (2001)             & Dragnet (1987)               & City of Violence (2006) \\
    \bottomrule
  \end{tabular}
  }
  
  \caption{\label{tab:hotshots}Results provided by our system for the
  movie \emph{Hot Shots!} (see Examples~\ref{ex:movielens},
  \ref{ex:tfidf}, and~\ref{ex:recommendation}).}
\end{table*}

\begin{example}\label{ex:recommendation}
  Consider the local utility election for the \emph{Hot Shots!}
  movie. In Table~\ref{tab:hotshots} we show the $p$-HUV committees
  for, $p \in \{0,1,2\}$, where for $0$-HUV we use the exact algorithm
  and for the other two rules we use simulated annealing and the
  greedy algorithm. Let us discuss the contents of these committees
  (for $p \in \{1,2\}$, we focus on simulated annealing):
  \begin{enumerate}
  \item The first six movies selected by $0$-HUV are quirky, absurd
    comedies, quite in spirit of \emph{Hot Shots!}. Among the next
    four movies, three are comedies (one of which is somewhat
    similar in spirit to the first six) and one is an action movie.

  \item Except for \emph{Yamakasi}, all movies selected by $1$-HUV are
    comedies of different styles, including four of the same style as
    \emph{Hot Shots!}, two action comedies, two family comedies, and
    one crime comedy. \emph{Yamakasi} is an action/drama movie which
    stands out from the rest.

  \item $2$-HUV selects even more varied set of comedies than $1$-HUV, including
    a western comedy, a sci-fi comedy, crime/action comedies, and family movies.
    Yet, it also includes \emph{Yamakasi}.
  \end{enumerate}

  The reason why \emph{Yamakasi} is included in our $1$-HUV and
  $2$-HUV committees is simply because, in total, it only received 53
  approvals, of which 27 came from people who enjoyed \emph{Hot
    Shots!}. Thus it was viewed as a very relevant movie for the
  query. If we replaced the simple \tfidf{} heuristic with a more
  involved scoring system (possibly using more information about the
  movies), we could account for such situations better.

  The committees computed by the greedy algorithm for $1$-HUV and
  $2$-HUV are of comparable quality to those provided by simulated
  annealing, although they include eight comedies and two action
  movies each.

\end{example}

\section{Experiments}

In this section we present four experiments. All but the second one
 are conducted on the MovieLens dataset, whereas the second one
uses synthetic data.  In the first experiment, we describe our process
of choosing the $\gamma$ value for the TF-IDF heuristic. In the second
one, 
we test if, indeed, $0$-HUV rule focuses on resources very similar to
the one from the query set, whereas $p$-HUV rules for
$p \in \{1,2,3\}$ seek increasingly more diverse result sets.
In the third experiment our goal is similar as in the first one, but we
analyze real-life data and we present a certain visualization of the
seach results.
In the
final experiment, we compare the performances of the greedy algorithm
and simulated annealing.

\subsection{Calibrating the TF-IDF Metric}\label{sec:startrek}

\begin{table}[b]
  \small\centering
  \begin{subtable}[b]{0.56\textwidth}
  \begin{tabular}{r|l}
    \toprule
         & Approval \\
   Movie & Count \\

    \midrule
                  Star Trek: Renegades  &      27   \\
                    Star Trek: Nemesis  &    1904   \\
                      Star Trek Beyond  &    1987   \\
       Star Trek V: The Final Frontier  &    2338   \\
               Star Trek: Insurrection  &    3783   \\
         Star Trek: The Motion Picture  &    3785   \\
   Star Trek III: The Search for Spock  &    4732   \\
Star Trek VI: The Undiscovered Country  &    5358   \\
               Star Trek Into Darkness  &    5621   \\
         Star Trek IV: The Voyage Home  &    7544   \\
       Star Trek II: The Wrath of Khan  &   10669   \\
                Star Trek: Generations  &   10809   \\
              Star Trek: First Contact  &   12396   \\
                             Star Trek  &   12854   \\
    \bottomrule
  \end{tabular}
    \caption{\label{tab:calib-movies}Star Trek movies present
      in the MovieLens 25M dataset.}
  \end{subtable}\quad\quad
  \begin{subtable}[b]{0.37\textwidth}
    \small\centering
  \begin{tabular}{r|l}
    \toprule
             & Average top \\
    $\gamma$ & ten count \\

    \midrule
    1.2 &   3.53 \\
    1.4 &   5.13 \\
    1.6 &   6.07 \\
    1.8 &   6.53 \\
    2.0 &   \textbf{6.73} \\
    2.2 &   6.47 \\
    2.4 &   5.73 \\
    2.6 &   3.93 \\
    2.8 &   1.80 \\
    \bottomrule
  \end{tabular}
  \caption{\label{tab:calib-top10-tfidf}Star Trek \tfidf{} calibration
    results.}

  \end{subtable}
  \caption{Summary of the Star Trek movies (left) and the results of
    calibrating the TF-IDF heuristic (right).}
\end{table}

\begin{figure*}
  \includegraphics[width=1\textwidth]{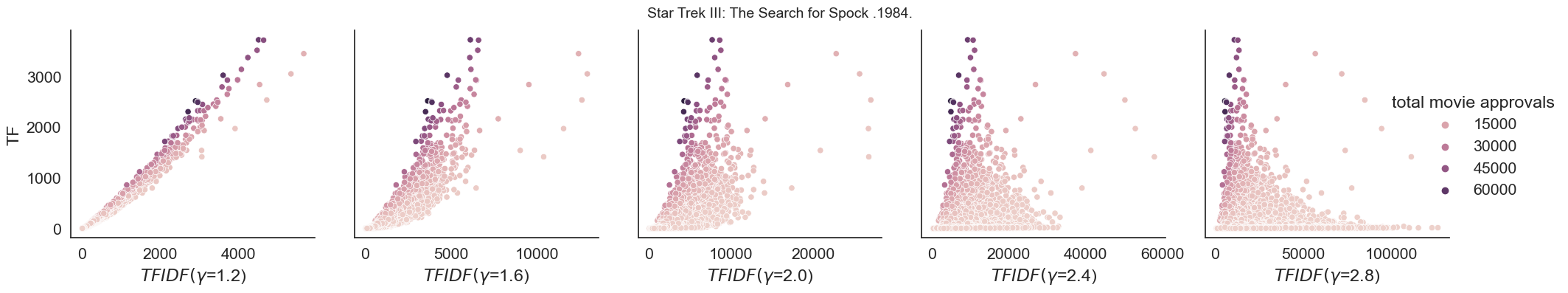}
  \caption{Each dot represents a movie in the local election generated
    for \emph{Star Trek III: The Search for Spock (1984)} and shows
    the relation between the movie's popularity (its \tf{} value) on
    the $\boldsymbol y$ axis and its final \tfidf{} score on the $\boldsymbol x$ axis for
    various setting of the parameter $\boldsymbol \gamma$. The hue of the dot
    represents the popularity of the movie in the global election
    (i.e., the number of its approvals)}
  \label{fig:calib-balance-tfidf_sfs}

  \includegraphics[width=1\textwidth]{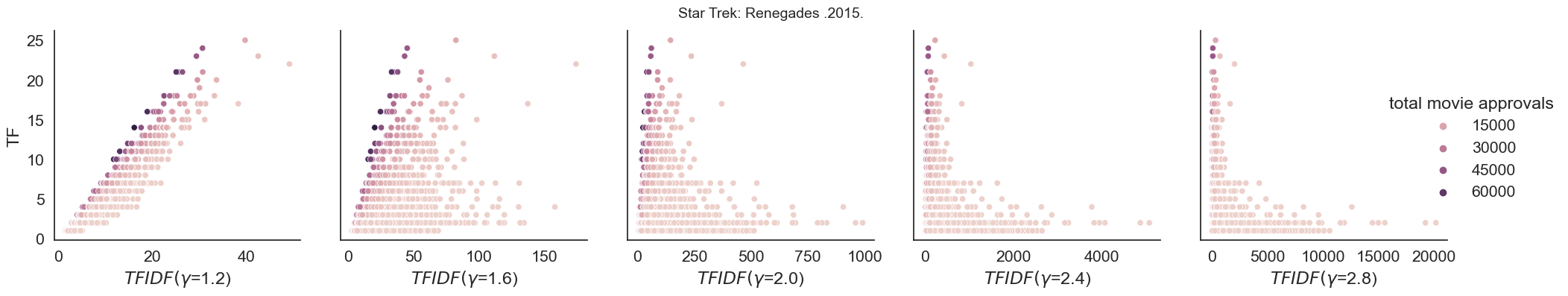}
\caption{Analogous to Figure~\ref{fig:calib-balance-tfidf_sfs} but for the
movie \emph{Star Trek: Renegades (2015)}.}
\label{fig:calib-balance-tfidf_renegades}

\end{figure*}

Before we describe our formal procedure for choosing the $\gamma$
parameter, let us explore its meaning. Intuitively, $\gamma$ is used
to give more weight to the IDF component relative to the TF one. In
other words, replacing $\gamma$ with a larger value more strongly
diminishes the TF-IDF values of the globally more popular movies than
of the less popular ones (i.e., those with fewer approvals in the
global election). This balance is visualised in
Figure~\ref{fig:calib-balance-tfidf_sfs}, where we consider the movie
\emph{Star Trek III: The Search for Spock (1984)} as the singleton
query, and for each movie in the local approval election we draw a dot
showing the relation between its number of approvals in the local
election (i.e., its TF value), on the $y$ axis, and its final TF-IDF
value, on the $x$ axis, for several values of $\gamma$. The hue of the
dot represents the number of approvals of the movie in the global
election. The top ten movies according to TF-IDF (for a given
$\gamma$) are the ten rightmost dots in the respective diagram. Note
that the higher the $\gamma$ is, the more dots with low TF value
appear to the right and have higher chance of being among top ten
movies. For $\gamma=1.2$, quite a few generally popular items (with
darker hue) make it to the top ten, simply because they are popular
overall and not only in the context of the search for a given query
set. For $\gamma=2.0$, there seems to be a good balance between the
popular and not so popular movies, while for $\gamma=2.8$ there are
only extremely unpopular movies selected for the top ten.

The above argument for using $\gamma = 2$ is based on intuition and to
get a better grounding for the choice, we have performed the following
experiment. Our basic premise is that the $\gamma$ value should be
such that when searching for a query set consisting of a single movie,
its most similar movies should appear among the top ten with respect
to the TF-IDF metric. While deciding what is ``the most similar movie''
is quite a subjective issue, we assumed that for all the movies
from the \emph{Star Trek} series, other \emph{Star Trek} movies are
the most similar ones. The MovieLens dataset contains fourteen
\emph{Star Trek} movies (that are approved by at least 20 users) and we
list them, together with their approval counts in the global election,
in Table~\ref{tab:calib-movies}.

\begin{figure*}
\includegraphics[width=1\textwidth]{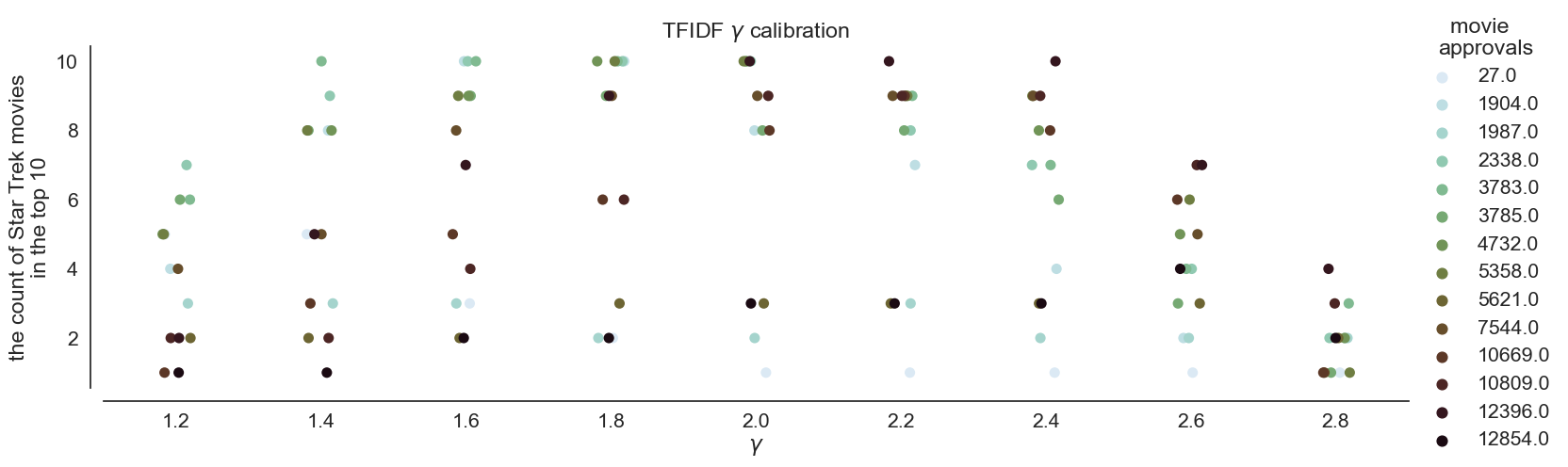}
\caption{The calibration experiment results. Each dot represents the \emph{top 10 count} value (vertical axis) when searching for a single movie (hue represents the popularity of the movie, the darker the more approvals the movie has) using \tfidf{} with a given $\gamma$ (horizontal axis).}
\label{fig:calib-top10-tfidf}
\end{figure*}

We used each \emph{Star Trek} movie as a singleton query set, computed
its local approval election, ranked the movies from this election with
respect to their TF-IDF values for $\gamma \in \{1.2, 1.4, 1.6,$
$1.8, 2.0, 2.2, 2.4, 2.6, 2.8\}$, and for each of these values of
$\gamma$ calculated how many other \emph{Star Trek} movies are among
the top ten ones.  We present the results in
Figure~\ref{fig:calib-top10-tfidf}. The interpretation of this figure
is as follows.  For each value of $\gamma$, we present all the 14
\emph{Star Trek} movies as dots. The $y$ coordinate of each dot is the
number of other \emph{Star Trek} movies that are among top-ten ones
when we use the given movie as a singleton query set. The color of
each dot corresponds to the number of approvals it receives in the
global election.  Finally, we have averaged the $y$ coordinates of the
fourteen movies for each value of $\gamma$ and present the results in
Table~\ref{tab:calib-top10-tfidf}. It turned out that $\gamma = 2$
indeed gave the best results (naturally, doing a more fine-grained
search for the value of $\gamma$ might lead to a slightly different
outcome, but we did not feel it would affect the other results in the
paper significantly).

While setting $\gamma = 2$ works well for many movies, it is not as
good a choice for some others.  Consider
Figure~\ref{fig:calib-balance-tfidf_renegades} which is analogous to
Figure~\ref{fig:calib-balance-tfidf_sfs}, but for  the
movie \emph{Star Trek: Renegades (2015)}, which is not very
popular and its local approval election contains relatively few movies
(dots on the diagram). Note that taking $\gamma = 2$ does not seem to
strike the right balance and we would need to choose $\gamma$ between
$1.2$ and $1.6$.  Nevertheless, we choose the simplicity of choosing a
single value of the $\gamma$ parameter for all the movies, at the cost
of getting a few of them wrong.

\subsection{Focus Versus Breadth: Synthetic Data}

\newcommand{\movie}[3]{#1.#2(#3)}

In the second experiment we generate the global election
synthetically, so that it resembles data regarding
movies.
The point is to observe the differences between committees computed
according to $p$-HUV rules for different values of $p$ in a controlled
environment.

\paragraph{Generating Global Elections.}
We assume that we have nine main categories of movies (such as, e.g.,
a comedy or a thriller) and each category has nine subcategories (such
as, e.g., a romantic comedy, or a psychological thriller). For each
pair of a category $u \in \{1, \ldots, 9\}$ and its subcategory
$v \in \{1, \ldots, 9\}$, we generate $25$ movies, denoted
$\movie{u}{v}{1}, \ldots, \movie{u}{v}{25}$.  Given a movie
$\movie{u}{v}{i}$, we set its \emph{quality factor} to be
$q(i) = -\arctan{\frac{i-13}{10}}+2$. That is, for each subcategory
the first movie has the highest quality, about $2.87$, and the
qualities of the following movies decrease fairly linearly, down to
about $1.12$.\footnote{The choice of this function is quite
  arbitrary. We could have used a purely linear one, or one that leads
  to a sharper difference between the top ``high-quality'' movies and
  bottom ``mediocre'' ones and the results would not change too
  much. Yet, changing the function, e.g., to an exponential one could
  have much stronger results.}
We also have $n = 2000$ voters.  Each voter $i$ has a probability
distribution $P_i$ over the main categories (where we interpret
$P_i(u)$ as the probability that the voter watches a movie from
category $u$), and for each category $u$, he or she has probability
distribution $P_i^{u}$ over its subcategories (interpreted as the
conditional probability that if the voter watches a movie from
category $u$, then this movie's subcategory is $v$). For each voter,
we choose these distributions as permutations of:
\[
  ( 0.5, 0.1, 0.1, 0.1, 0.1, 0.025, 0.025, 0.025, 0.025 )
\]
chosen uniformly at random.
Intuitively, each voter has his or her most preferred category, four
categories that he or she also quite enjoys, and four categories that
he or she rarely enjoys (the same applies to subcategories).  To
generate an approval of a voter $i$ we do as follows:
\begin{enumerate}
\item We choose a category $u$ according to distribution $P_i$ and,
  then, a subcategory $v$ according to distribution $P_i^u$.
\item We choose a movie among
  $\movie{u}{v}{1}, \ldots, \movie{u}{v}{25}$ with probability
  proportional to its quality factor. The voter approves the selected
  movie.
\end{enumerate}
We repeat this process 162 times for each voter, leading to a bit
fewer approvals (due to repetitions in sampling; recall that in
MovieLens the average number of approvals is 154).

While the above-described process of generating a global election is
certainly quite ad-hoc, we believe that it captures some of the main
features of people's preferences regarding movies. More importantly,
we can say that two movies are very similar if they come from the same
subcategory, are somewhat similar if they come from the same category
but different subcategories, and are very loosely related otherwise.

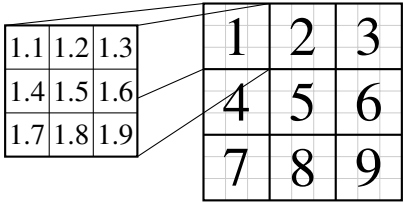
\begin{wrapfigure}{l}{7cm}
  \centering
  \begin{tikzpicture}[scale=0.29]
    \draw[step=1,gray!40,very thin] (0,0) grid (9,9);

    \draw[thick] (0,0) -- (9,0) -- (9,9) -- (0,9) -- cycle;
    \draw[thick] (3,0) -- (3,9);
    \draw[thick] (6,0) -- (6,9);
    \draw[thick] (0,3) -- (9,3);
    \draw[thick] (0,6) -- (9,6);

    \foreach \u in {1,2,3}
       \draw (1.5+3*\u-3,6+1.5) node {\huge \u};
    \foreach \u in {4,5,6}
       \draw (1.5+3*\u-12,3+1.5) node {\huge \u};
    \foreach \u in {7,8,9}
        \draw (1.5+3*\u-21,0+1.5) node {\huge \u};

    \draw (-9,8) -- (0,9);
    \draw (-9,2) -- (0,6);    
    \filldraw[thick, fill=white] (-9,2) -- (-3,2) -- (-3,8) -- (-9,8) -- cycle;
    \draw (-3,2) -- (3,6);
    \draw (-3,8) -- (3,9);

    \draw (-7,2) -- (-7,8);
    \draw (-5,2) -- (-5,8);
    \draw (-9,4) -- (-3,4);
    \draw (-9,6) -- (-3,6);

    \foreach \u in {1,2,3}
       \draw (-9+1+2*\u-2,6+1) node {1.\u};
    \foreach \u in {4,5,6}
       \draw (-9+1+2*\u-8,4+1) node {1.\u};
    \foreach \u in {7,8,9}
       \draw (-9+1+2*\u-14,2+1) node {1.\u};
    
  \end{tikzpicture}
  \caption{\label{fig:categories}Visual arrangement of the movie
    categories and subcategories for the synthetic experiment.}
  \vspace{-0.2cm}
\end{wrapfigure}

\paragraph{Running The Experiment.}

For each number $p \in \{0, 1, 2, 3\}$ and both algorithms for
computing approximate $p$-HUV committee (i.e., the greedy algorithm
and simulated annealing) we repeat the following
experiment.\footnote{We compute the $0$-HUV committees using the
  optimal, polynomial-time algorithm.} We generate $100$ global
elections as described above, and for each of them we compute a
committee of size $k=10$ for the query set consisting of movie
$\movie{1}{1}{13}$, i.e., the middle-quality movie from subcategory
$1.1$ (since the (sub)categories are, effectively, symmetric, their
choice is irrelevant). Altogether, $1000$ movies are selected (some of
the movies are selected more than once and we count each of
them). Then, for each subcategory $u.v$, we sum up, over all the
computed committees, how many movies from this subcategory were
selected, obtaining a histogram (in this respect, our experiment is
quite similar to that of
\citet{elk-fal-las-sko-sli-tal:c:2d-multiwinner}).

\begin{figure*}[t]
  \centering
  \begin{subfigure}[b]{3cm}
    \centering
    \begin{tikzpicture}[scale=1.5]
      \draw (0,0) node {\includegraphics[width=3cm]{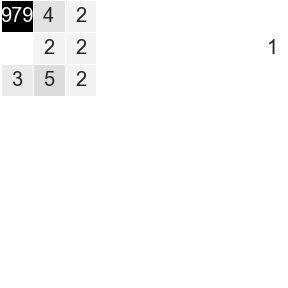}};
      \draw[thick] (-1,0-1/3) -- (1,0-1/3);
      \draw[thick] (-1,0+1/3) -- (1,0+1/3);
      \draw[thick] (0-1/3,1) -- (0-1/3,-1);
      \draw[thick] (0+1/3,1) -- (0+1/3,-1);
      \draw[thick] (-1,-1) -- (-1,1) -- (1,1) -- (1,-1) -- (-1,-1);
    \end{tikzpicture}
         \caption{\small \centering $0$-HUV,  ${(979,20,1)}$}
         \label{fig:sa:0}
  \end{subfigure}\quad
  \begin{subfigure}[b]{3cm}
    \centering
    \begin{tikzpicture}[scale=1.5]
      \draw (0,0) node {\includegraphics[width=3cm]{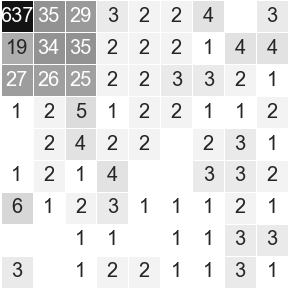}};
      \draw[thick] (-1,0-1/3) -- (1,0-1/3);
      \draw[thick] (-1,0+1/3) -- (1,0+1/3);
      \draw[thick] (0-1/3,1) -- (0-1/3,-1);
      \draw[thick] (0+1/3,1) -- (0+1/3,-1);
      \draw[thick] (-1,-1) -- (-1,1) -- (1,1) -- (1,-1) -- (-1,-1);
    \end{tikzpicture}
         \caption{\small \centering $1$-HUV, ${(637,230,133)}$}
         \label{fig:sa:1}
  \end{subfigure}\quad
  \begin{subfigure}[b]{3cm}
    \centering
    \begin{tikzpicture}[scale=1.5]
      \draw (0,0) node {\includegraphics[width=3cm]{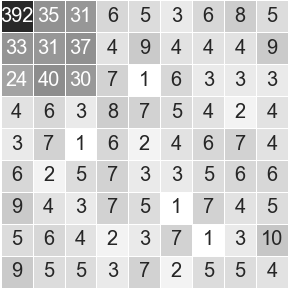}};
      \draw[thick] (-1,0-1/3) -- (1,0-1/3);
      \draw[thick] (-1,0+1/3) -- (1,0+1/3);
      \draw[thick] (0-1/3,1) -- (0-1/3,-1);
      \draw[thick] (0+1/3,1) -- (0+1/3,-1);
      \draw[thick] (-1,-1) -- (-1,1) -- (1,1) -- (1,-1) -- (-1,-1);
    \end{tikzpicture}
         \caption{\small \centering $2$-HUV, ${(392,261,347)}$}
         \label{fig:sa:2}
  \end{subfigure}\quad
  \begin{subfigure}[b]{3cm}
    \centering
    \begin{tikzpicture}[scale=1.5]
      \draw (0,0) node {\includegraphics[width=3cm]{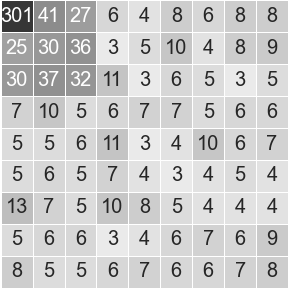}};
      \draw[thick] (-1,0-1/3) -- (1,0-1/3);
      \draw[thick] (-1,0+1/3) -- (1,0+1/3);
      \draw[thick] (0-1/3,1) -- (0-1/3,-1);
      \draw[thick] (0+1/3,1) -- (0+1/3,-1);
      \draw[thick] (-1,-1) -- (-1,1) -- (1,1) -- (1,-1) -- (-1,-1);
    \end{tikzpicture}
         \caption{\small \centering $3$-HUV,  ${(301,258,441)}$}
         \label{fig:sa:3}
  \end{subfigure}
  \caption{\label{fig:sa}Histograms for the synthetic experiment and simulated annealing.}\vspace{0.5cm}

  \centering
  \begin{subfigure}[b]{3cm}
    \centering
    \begin{tikzpicture}[scale=1.5]
      \draw (0,0) node {\includegraphics[width=3cm]{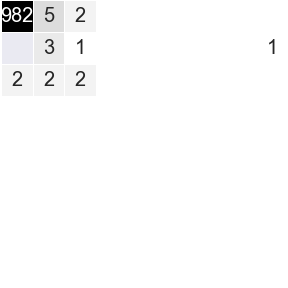}};
      \draw[thick] (-1,0-1/3) -- (1,0-1/3);
      \draw[thick] (-1,0+1/3) -- (1,0+1/3);
      \draw[thick] (0-1/3,1) -- (0-1/3,-1);
      \draw[thick] (0+1/3,1) -- (0+1/3,-1);
      \draw[thick] (-1,-1) -- (-1,1) -- (1,1) -- (1,-1) -- (-1,-1);
    \end{tikzpicture}
         \caption{\small \centering $0$-HUV, ${(982,17,1)}$}
         \label{fig:greedy:0}
  \end{subfigure}\quad
  \begin{subfigure}[b]{3cm}
    \centering
    \begin{tikzpicture}[scale=1.5]
      \draw (0,0) node {\includegraphics[width=3cm]{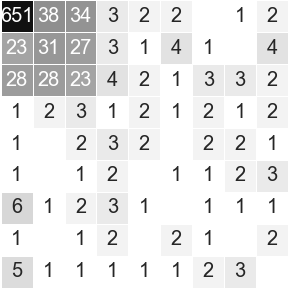}};
      \draw[thick] (-1,0-1/3) -- (1,0-1/3);
      \draw[thick] (-1,0+1/3) -- (1,0+1/3);
      \draw[thick] (0-1/3,1) -- (0-1/3,-1);
      \draw[thick] (0+1/3,1) -- (0+1/3,-1);
      \draw[thick] (-1,-1) -- (-1,1) -- (1,1) -- (1,-1) -- (-1,-1);
    \end{tikzpicture}
         \caption{\small \centering $1$-HUV, ${(651,232,117)}$}
         \label{fig:greedy:1}
  \end{subfigure}\quad
  \begin{subfigure}[b]{3cm}
    \centering
    \begin{tikzpicture}[scale=1.5]
      \draw (0,0) node {\includegraphics[width=3cm]{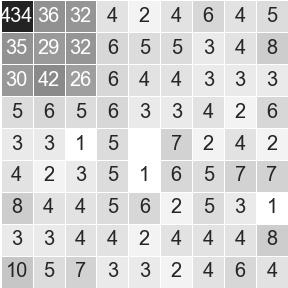}};
      \draw[thick] (-1,0-1/3) -- (1,0-1/3);
      \draw[thick] (-1,0+1/3) -- (1,0+1/3);
      \draw[thick] (0-1/3,1) -- (0-1/3,-1);
      \draw[thick] (0+1/3,1) -- (0+1/3,-1);
      \draw[thick] (-1,-1) -- (-1,1) -- (1,1) -- (1,-1) -- (-1,-1);
    \end{tikzpicture}
         \caption{\small \centering $2$-HUV, ${(434,262,304)}$}
         \label{fig:greedy:2}
  \end{subfigure}\quad
  \begin{subfigure}[b]{3cm}
    \centering
    \begin{tikzpicture}[scale=1.5]
      \draw (0,0) node {\includegraphics[width=3cm]{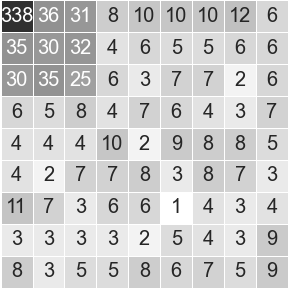}};
      \draw[thick] (-1,0-1/3) -- (1,0-1/3);
      \draw[thick] (-1,0+1/3) -- (1,0+1/3);
      \draw[thick] (0-1/3,1) -- (0-1/3,-1);
      \draw[thick] (0+1/3,1) -- (0+1/3,-1);
      \draw[thick] (-1,-1) -- (-1,1) -- (1,1) -- (1,-1) -- (-1,-1);
    \end{tikzpicture}
         \caption{\small \centering $3$-HUV, ${(338,254,408)}$}
         \label{fig:greedy:3}
  \end{subfigure}
  \caption{\label{fig:greedy}Histograms for the synthetic experiment and the greedy algorithm.}
\end{figure*}

To present these histograms visually, we arrange the categories into a
$3 \times 3$ square, where each category is further represented as a
$3 \times 3$ subsquare of subcategories, as shown in
Figure~\ref{fig:categories}.  We show thus-arranged histograms for
simulated annealing in Figure~\ref{fig:sa} and for the greedy
algorithm in Figure~\ref{fig:greedy}. Each subcategory square is
labeled with the number of movies selected from this subcategory and
its background reflects this number (darker backgrounds correspond to
higher numbers).
Further, in both figures next to the name of each $p$-HUV rule we
report a vector $(x,y,z)$, where $x$ means the number of movies
selected from subcategory $1.1$, $y$ means the number of movies
selected from category $1$ except for those in subcategory $1.1$, and
$z$ refers to the number of all the other selected movies.  Thus we always
have that $x+y+z = 1000$.

\paragraph{Analysis.}
Our main conclusion is that, indeed,
$0$-HUV focuses on very similar movies (almost all the selected movies
come from category $1.1$) and as $p$ increases, approximate $p$-HUV
committees include more and more movies from other subcategories of
category $1$, and, eventually, even more movies outside of it.
It would be desirable to have a value of $p$ for which we would get a
vector $(x,y,z)$ close to, say, $(450, 450, 100)$, meaning that, on
average, the resulting committee would contain between $4$ and $5$
movies from category $1.1$ (i.e., directly relevant to the search),
between $4$ and $5$ movies from other subcategories of category $1$
(i.e., similar but quite different from the query), and $1$ movie from
some other category (i.e., something very different, but possibly
appealing to the people who enjoyed the movie from the query). Yet,
our algorithms do not seem to provide committees with such vectors
(this, however, is not a major worry---after all, the setup in the
experiment was simplified and, to some extent,
Example~\ref{ex:recommendation} shows that for real-life MovieLens
data we do find such committees).

\subsection{Focus Versus Breadth: Movies Data}

In this section we observe how varying the value of $p$ affects the
results of $p$-HUV rules on the MovieLens dataset. Ideally, we would
like to see the same transition from very focused to quite broad
results as with synthetic data. To this end, we first need a
methodology for analyzing similarity between movies.
We focus on the greedy algorithm for computing $p$-HUV
rules as the results for simulated annealing would be very similar.

\paragraph{Similarity Among Movies.}
Let us consider two movies, $x$ and $y$. We take the following
approach to obtain a number that, in some way, is related to their
similarity. First, we form a local utility election using~$x$ as the
singleton query set (to maintain symmetry, later we will do the same
for~$y$). If~$y$ does not appear in this election, then we consider it
as completely dissimilar from $x$. Otherwise, we sort the movies from
the local election in the ascending order of their TF-IDF values (as
before, we use $\gamma = 2$) and we define the rank of $y$ with
respect to $x$, denoted $\rank_x(y)$ to be the position on which $y$
appears. (So if $y$ has the highest TF-IDF value then
$\rank_x(y) = 1$, if it has the second highest TF-IDF then
$\rank_y(x)= 2$, and so on; recall that the idea of TF-IDF is that the
higher it is, the more relavant a movie is for the search query and,
so, we equate relevance with similarity). We define the dissimilarity
between $x$ and $y$ as:
\[
  \dis(x,y) = \nicefrac{1}{2}\big( \rank_x(y) + \rank_y(x) \big).
\]
This definition ensures that $\dis(x,y) = \dis(y,x)$ and that the
larger $\dis(x,y)$ is, the less related---and, hence, less
similar---are the two movies.  However, it also has some flaws. For
example, let us consider two sets of movies, $A$ and $B$, where the
movies in each of the sets are very similar to each other. Further,
let us assume that $A$ contains significantly more movies than $B$.
Then the average dissimilarity between the movies in $A$ would be
considerably larger than the dissimilarity between the movies in $B$,
even if objectively there would be no justification for such a
difference. While we acknowledge that such issues may happen, we view
them as somewhat extreme and we expect that in most situations our
measure is sufficient to distinguish between movies that are clearly
related and those that do not have much to do with each other.

\paragraph{Gathering Movies for Comparison.}

We would like to compare the outcomes of various $p$-HUV rules on each
of the movies from some set $A$. To do so, we form an extension of $A$
as follows (we consider~$p$ values in $\{0,1,2,3\}$ and committee size
$k=10$, unless we say otherwise):
\begin{enumerate}
\item For each movie $x \in A$ and each $p \in \{0,1,2,3\}$, we
  compute the $p$-HUV winning committee for the local utility election
  based on the singleton query $x$. We take the union of these
  committees and call it~$B$.

\item For each movie $y \in B$, we compute $0$-HUV winning committee
  of size two for the local utility election based on $y$. We refer to
  the union of these committees as~$C$.

\item We declare $\ext(A) = A \cup B \cup C$ to be the extension of $A$.
\end{enumerate}
We use set $B$ because we are interested in relations between the
contents of the committees provided by all our rules for all the
movies in $A$, and we add set $C$ because we also want to ensure that
for each member of one of the committees there also are some very
similar movies in the extension.

\paragraph{Visualizing Relations Between the Movies.}
Given a set $A$ of movies and its extension $\ext(A)$, we first
compute the value $\diss(x,y)$ for all distinct movies $x$ and $y$ in
$\ext(A)$. Then, we form a complete graph where members of $\ext(A)$
are the nodes and for each two movies $x$ and $y$, the edge connecting
them has weight $\diss(x,y)$. Then we compute an embedding that maps
each movie in $\ext(A)$ to a point on a two-dimensional plane, so that
the Euclidean distances between these points correspond to the weights
of the edges. To this end, we use the force-directed algorithm of
\citet{fru-rei:j:force-directed}. We use an implementation of the
algorithm as provided in the \emph{networkx} library
(\emph{networkx.drawing.layout.spring\_layout}, version $2.6.3$), ran
for 10'000 iterations. For a description of the library we refer the
reader to \cite{hagberg2008exploring}.

As input the Fruchterman-Reingold algorithm does not take the weights
of the edges in the graph whose embedding it is to compute, but the
forces that act to bring the nodes of a given edge closer. Since this
value should be inverse to our dissimilarity, for each two movies $x$
and $y$ we use force $(\diss(x,y))^{-2}$ (i.e., we use the square of
the inversed dissimilarity; by experimenting with different force
functions we found this value to work well).

\paragraph{Experiment 1: Comparing $\boldsymbol{p}$-HUV Rules.}

\begin{figure*}
  \centering
  \includegraphics[width=11cm]{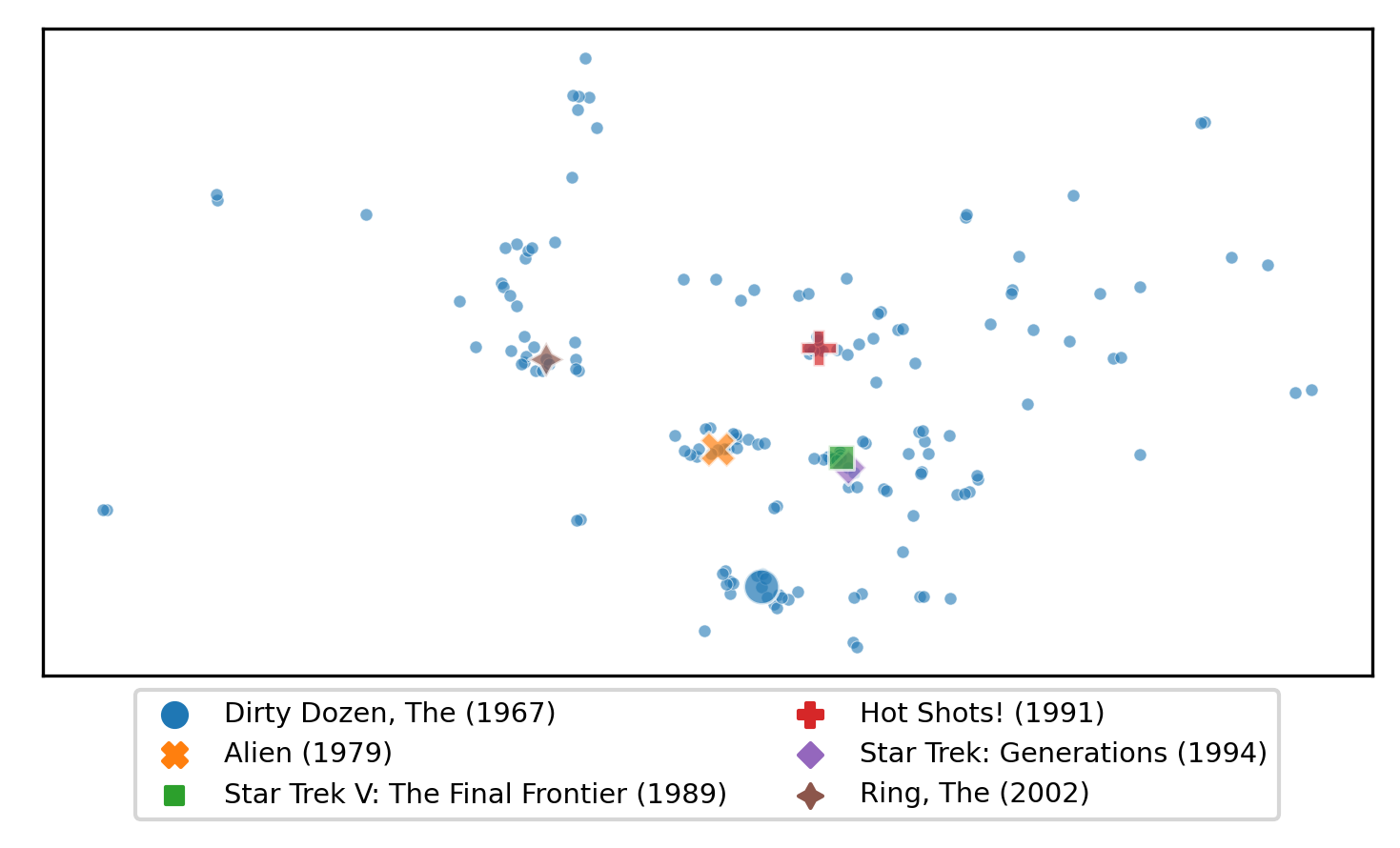}
  \caption{\label{fig:classic-map}Embedding of the movies in Experiment~1.}
\end{figure*}

Next we use the visualization methodology to analyze the outcomes of
different $p$-HUV rules. We let the set $A$ consist of the following
movies: \emph{Hot Shots! (1991)}, \emph{Ring, The (2002)}, \emph{Star
  Trek V: The Final Frontier (1989)}, \emph{Star Trek: Generations
  (1994)}, \emph{Alien (1979)}, and \emph{Dirty Dozen, The (1967)}.
We chose these movies because they represent different genres and
styles.  \emph{Hot Shots!} is a quirky comedy, \emph{The Ring} is a
horror movie, \emph{Star Trek} movies are examples of science-fiction,
and so is \emph{Alien} which also has strong elements of a horror
movie. Finally, \emph{The Dirty Dozen} is a classic war movie. We show
an embedding of these movies (and the movies from their extension) in
Figure~\ref{fig:classic-map}. In particular, we see that the two
\emph{Star Trek} movies are correctly presented as very similar,
whereas the other movies are farther away from each other.

\begin{figure*}
  \centering
  \includegraphics[width=16cm]{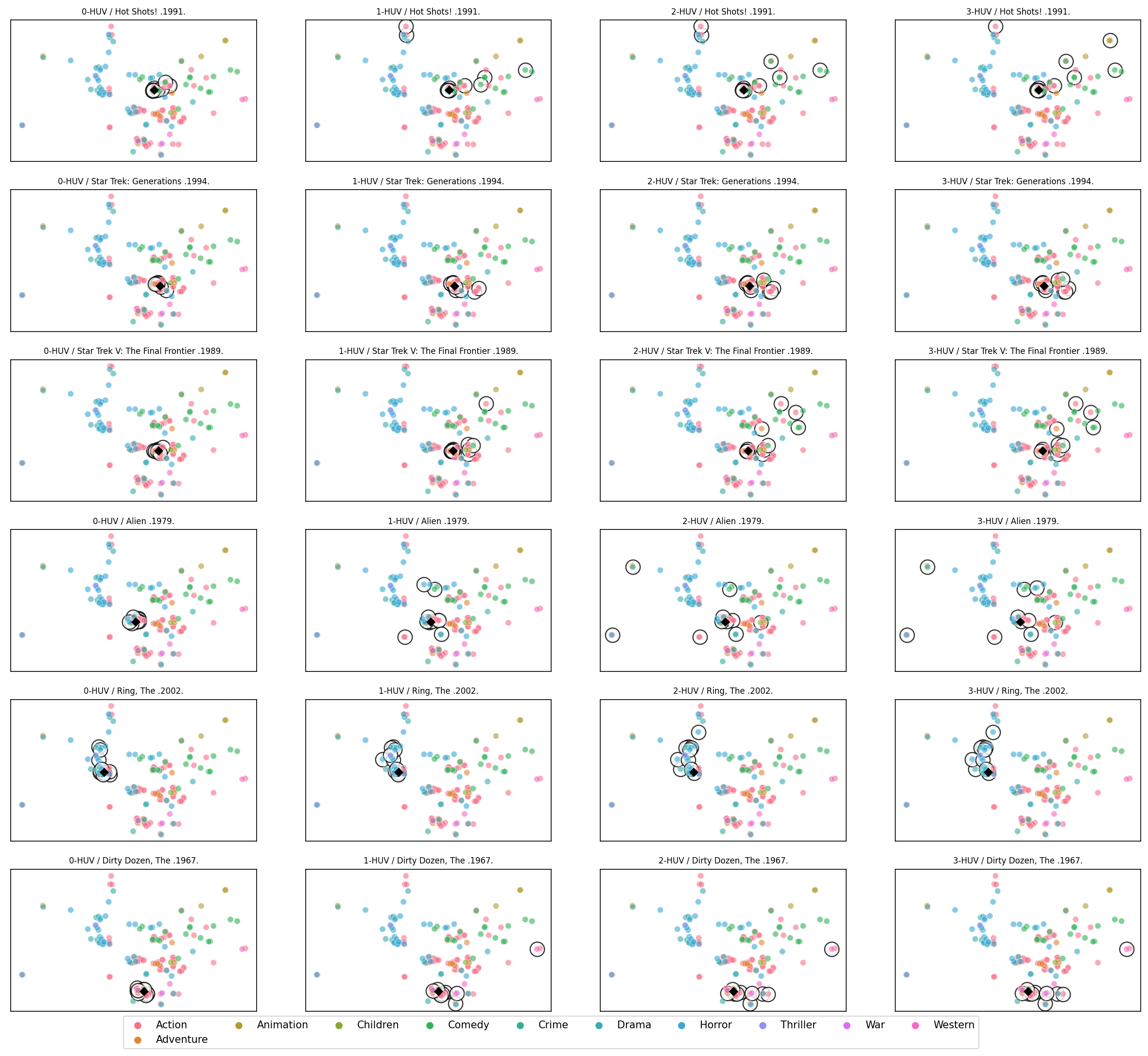}
  \caption{Committees computed using $p$-HUV rules for different queries from Experiment 1.}
  \label{fig:classic-dist}
\end{figure*}

In Figure~\ref{fig:classic-dist} we visualize the committees provided
by $p$-HUV rules for $p \in \{0,1,2,3\}$ and singleton query sets from
$A$. Specifically, each column corresponds to a value of $p$ and each
row to a different query. The members of selected committees have
surrounding black circles. The movies are colored with respect to the
first genre provided for it in MovieLens (in this dataset each movie
can have multiple genres, listed in some order).

We make several observations. First, we see that the $0$-HUV rule
always chooses movies very close (very similar) to the one from the
query. This is hardly surprising as, in essence, we have defined our
dissimilarity function to encode this effect. Thus it is more
interesting to consider $p$-HUV rules for $p \geq 1$. In this case, we
see that the selected movies are always farther away from the query
than for $0$-HUV, but the extent to which this happens varies.  For
example, for \emph{Hot Shots!} the committees get more and more spread
as $p$ increases, whereas for \emph{Star Trek: Generations} the
outcomes are very similar for all $p \geq 1$. It is also quite
striking that there is very little difference between $2$-HUV and
$3$-HUV for all the six movies. Yet, altogether, we can say that our
system achieves its main goals. The $0$-HUV rule gives tightly focused
results, closely connected to the query, and $p$-HUV rules for
$p \geq 1$ give more diverse results.

\paragraph{Experiment 2: Indiana Jones and Star Trek Movies}

\begin{figure*}
  \centering
  \begin{subfigure}[b]{7cm}
    \includegraphics[width=7cm]{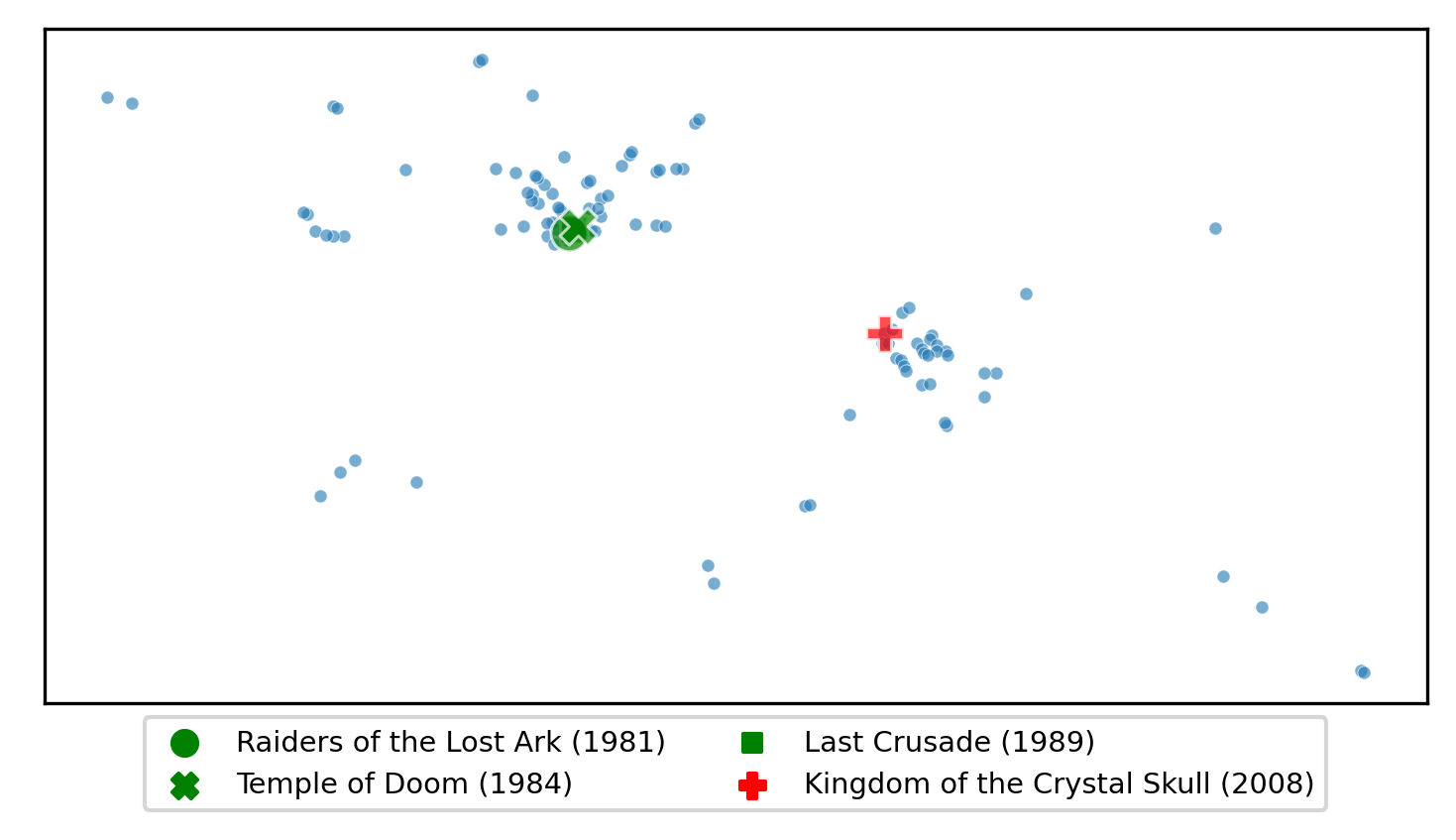}

    \rule{0cm}{0.98cm}
    \caption{\label{fig:indiana-map}Indiana Jones movies.}
  \end{subfigure}
  \begin{subfigure}[b]{7cm}
    \includegraphics[width=7cm]{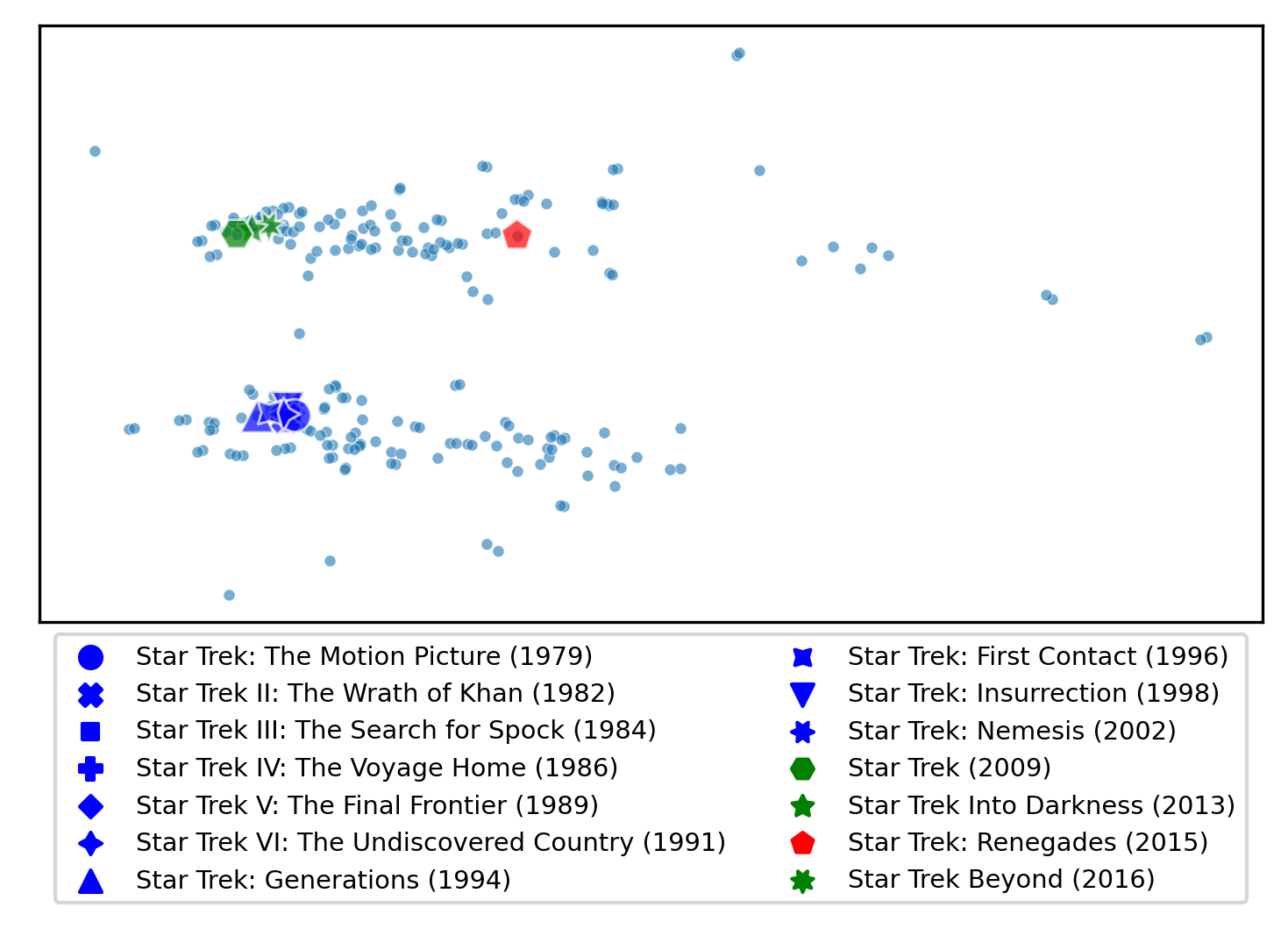}
    \caption{\label{fig:startrek-map}Star Trek movies.}
  \end{subfigure}
  \caption{Embeddings for Experiment~2.}
\end{figure*}

The second experiment provides a somewhat anectodical evidence that
our dissimilarity measure can identify interesting features of our
data. In this case, we first let $A$ consists of four movies about
Indiana Jones, an adventurer arachaeologist. These movies are:
\emph{Indiana Jones and the Raiders of the Lost Ark (1981)},
\emph{Indiana Jones and the Temple of Doom~(1984)}, \emph{Indiana
  Jones and the Last Crusade (1989)}, \emph{Indiana Jones and the
  Kingdom of the Crystal Skull (2008)}. We show the visualization of
the extension of this set of movies in Figure~\ref{fig:indiana-map}.

We see that the original trilogy, filmed in the 80s, is placed close
together, whereas the new movie from 2008 is quite far away. Indeed,
as one may expect, reviving of Indiana Jones franchise after nineteen
years resulted in very mixed reactions from the fans. We find it quite
interesting that such phenomena are visible in our dissimilarity
measure.
Yet, the readers may wonder if the difference between the \emph{Indiana
  Jones} movies can simply be explained by the difference in their
release dates. After all, it is quite natural that much older movies
might be approved by different sets of people than newer ones. This,
of course, might be the case, but it certainly is not the only
explanation.

To see that the release date may not influence similarity as much, let
us consider the \emph{Star Trek} movies, as listed in
Table~\ref{tab:calib-movies}, which were released between 1979 and
2016.  We show their embedding in Figure~\ref{fig:startrek-map}.  We
see that the first ten movies, released between 1979 and 2002 (i.e.,
over a period of over twenty years) and marked with blue symbols, are
clustered closely together. These movies form the original series
together with \emph{The Next Generation} series (the transition
between the two series was quite gentle, hence it is not surprising
that the two groups of movies come together\footnote{The transition
  happned over the 1991 and 1994 movies. Both were featuring the cast
  from the original series and from \emph{The Next Generation} one,
  with the 1991 movie focused on the former and the 1994 one focused
  on the latter.}). The next cluster consists of three movies from
2009, 2013, and 2016 and is marked in green. These movies form a new,
reboot series (so-called \emph{Kelvin Timeline}). Finally, the 2015
movie is a fan film and does not belong to the official set of movies.
Altogether, we see that similar movies are grouped together even if
they were released over a long period of time, whereas significant
changes, such as making a reboot or shooting a fan film, are clearly
separated.

\subsection{Effectiveness of the Approximations}

\begin{figure*}
  \centering
  \begin{subfigure}[b]{0.65\textwidth}
  \centering
\includegraphics[width=1\textwidth]{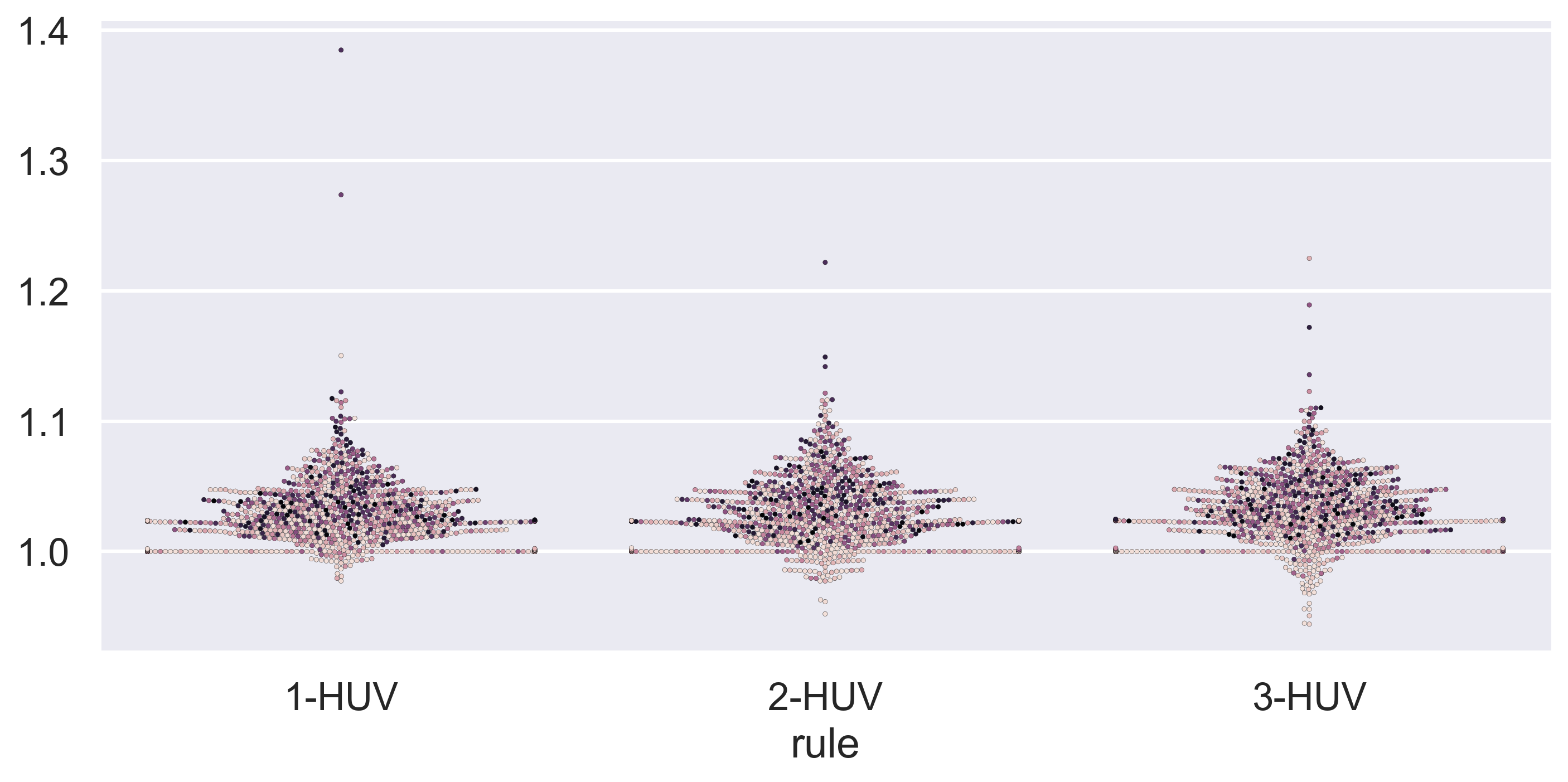}
\caption{Effectiveness of the greedy algorithm versus simulated
  annealing. 
  Each dot represents a single movie (from a set of 1000
  randomly selected ones), and is used as a singleton query set. Its
  position on the $y$ axis is the ratio of the scores of the
  committees computed for this movie using the greedy algorithm and
  simulated annealing. The position on the $x$-axis is perturbed to
  show all the dots. The color of the dot represents the number of
  approvals of the movie in the global election (the darker it is, the
  more approvals).}
\label{fig:run25m-fig-swarm}
  \end{subfigure}\quad
  \begin{subfigure}[b]{0.3\textwidth}
    \centering
      \begin{tabular}{r|ll}
        \toprule
        rule    &   Mean    &   Std. Dev. \\
        \midrule
        \input{res/algo_quality/basic_1k/scores-aggregated.tex}
        \bottomrule
        \multicolumn{3}{}{}
        \vspace{0.3cm}
      \end{tabular}
    \caption{\label{tab:run25m-tab}Average ratios of committee scores
    computed using the greedy algorithm and using the simulated annealing algorithm. We provide both the mean values of the scores and their standard deviations.\\}
  \end{subfigure}
  \caption{Effectiveness of our algorithms of the MovieLens 25M dataset.}
\end{figure*}

In the final experiment we compare the quality of the committees
computed by the greedy algorithm and by simulated annealing, for the
MovieLens dataset.
To this end, we sampled $1000$ movies and used them as singleton query
sets.  For each we have computed the local utility election computed
approximate winning committees for $1$-HUV, $2$-HUV, and $3$-HUV,
using both our algorithms. For each movie we calculated the ratio of
the score computed using the greedy algorithm and simulated annealing
(so if the ratio is above $1$ then the greedy algorithm performs
better and if it is below $1$, then simulated annealring is
better). We show the results in Figure~\ref{fig:run25m-fig-swarm} and,
in a more aggregate form, in Figure~\ref{tab:run25m-tab}.  On average,
the greedy algorithm finds committees with about $3\%$ higher scores
than simulated annealing.  Yet, as we have seen in
Example~\ref{ex:recommendation}, simulated annealing has other
positive features.  We note that we ran the simulated annealing
algorithm for $50 000$ steps and using more would certainly improve
the result (yet, simulated annealing already takes four times as long
to compute as the greedy algorithm).  All in all, we conclude that
both algorithms perform comparably well.

\section{Conclusions}

We have shown that multiwinner voting can be successfully used to
build a system that helps searching for movies and that lets the users
specify how strongly related should the proposed set of movies be to
those he or she asks about.  Interestingly, our approach can be used
to identify nonobvious relations between the movies (as we have
observed for the case of \emph{Indiana Jones} and \emph{Star Trek}
series of movies).  Our system does not use any advanced tools for
non-personalized recommendation systems and purely demonstrates that
multiwinner voting is something that designers of such systems might
want to consider.

We do not discuss running times of our algorithms. They are
implemented using Python and are highly non-optimized, so analyzing
their running times would not be very meaningful.

\section*{Acknowledgments}
  Grzegorz Gawron was supported in part by AGH University of Science
  and Technology and the "Doktorat Wdro\.zeniowy" program of the
  Polish Ministry of Science and Higher Education.
  This project has received funding
  from the European Research Council (ERC) under the European Union’s
  Horizon 2020 research and innovation programme (grant agreement No
  101002854).

\begin{flushright}  
    \includegraphics[width=4cm]{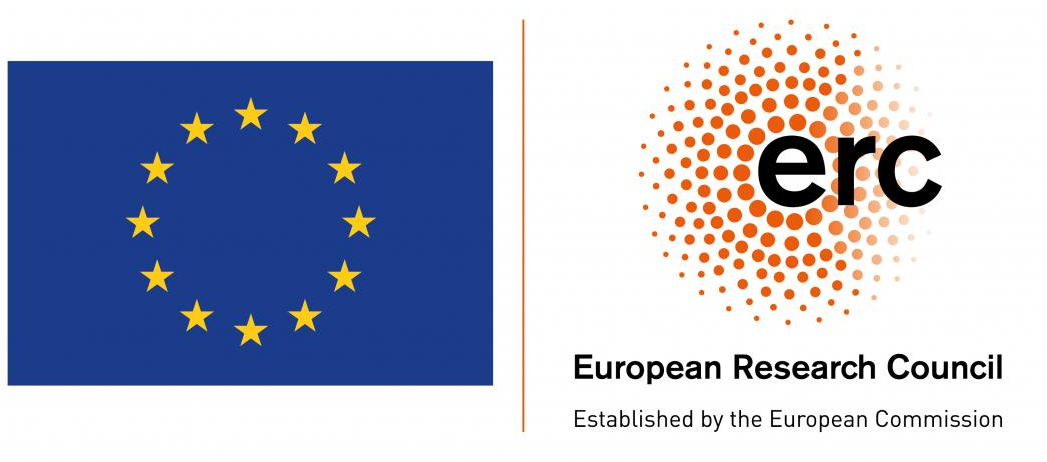}
\end{flushright} 

\bibliographystyle{ACM-Reference-Format}  
\bibliography{bib-using}

\end{document}